\documentclass [superscriptaddress,reprint,amsmath,amssymb,aps,prl,showkeys,floatfix]{revtex4-2}
\usepackage{xcolor}
\usepackage{soul}
\usepackage[normalem]{ulem}
\usepackage[hidelinks]{hyperref}
\usepackage{graphicx}
\usepackage{dcolumn}
\usepackage{bm}
\usepackage{color}
\usepackage{float}
\usepackage[utf8]{inputenc}
\usepackage[mathlines]{lineno}
\usepackage{multirow}
\usepackage{amsmath}
\usepackage{pdfpages}
\usepackage{pgffor}
\makeatletter
\AtBeginDocument{\let\LS@rot\@undefined}
\makeatother
\usepackage{hyperref}
\usepackage{soul,xcolor}
\usepackage{natbib}
\usepackage{amsmath}
\usepackage{braket}
\usepackage{xcolor}
\usepackage{xr}
\usepackage{chemformula}
\usepackage{siunitx}

\soulregister\cite7
\soulregister\ref7
\setstcolor{blue}
\newcommand{\doHMN}[2]{%
  \begingroup\lccode`~=`#1
  \lowercase{\endgroup\let~}#2%
  \mathcode`#1="8000
}

\makeatletter
\newcommand*{\addFileDependency}[1]{
  \typeout{(#1)}
  \@addtofilelist{#1}
  \IfFileExists{#1}{}{\typeout{No file #1.}}
}
\makeatother

\usepackage{tikz}

\begin{document}
\preprint{APS/123-QED}
\title{Poincar\'{e} sphere engineering of dynamical ferroelectric topological solitons}
\author{Lingyuan Gao}
\thanks{These authors contributed equally}
\affiliation{Smart Ferroic Materials Center, Physics Department and Institute for Nanoscience and Engineering, University of Arkansas, Fayetteville, Arkansas, 72701, USA}
\author{Yijie Shen}
\thanks{These authors contributed equally}
\email{yijie.shen@ntu.edu.sg}
\affiliation{Centre for Disruptive Photonic Technologies, School of Physical and Mathematical Sciences, Nanyang Technological University, Singapore 637371, Singapore}
\affiliation{School of Electrical and Electronic Engineering, Nanyang Technological University, Singapore 639798, Singapore}
\author{Sergei Prokhorenko}
\affiliation{Smart Ferroic Materials Center, Physics Department and Institute for Nanoscience and Engineering, University of Arkansas, Fayetteville, Arkansas, 72701, USA}
\author{Yousra Nahas}
\affiliation{Smart Ferroic Materials Center, Physics Department and Institute for Nanoscience and Engineering, University of Arkansas, Fayetteville, Arkansas, 72701, USA}
\author{Laurent Bellaiche}
\email{laurent@uark.edu}
\affiliation{Smart Ferroic Materials Center, Physics Department and Institute for Nanoscience and Engineering, University of Arkansas, Fayetteville, Arkansas, 72701, USA}
\affiliation{Department of Materials Science and Engineering, Tel Aviv University, Ramat Aviv, Tel Aviv 6997801, Israel}

\date{\today}

\begin{abstract}
Geometric representation lays the basis for understanding and flexible tuning of topological transitions in many physical systems. An example is given by the Poincar\'{e} sphere (PS) that provides an intuitive and continuous parameterization of the spin or orbital angular momentum (OAM) light states. Here, we apply this geometric construction to understand and continuously encode dynamical topologies of ferroelectric solitons driven by OAM-tunable light.
We show that: (1) PS engineering enables controlled creation of dynamic polar antiskyrmions that are rarely found in ferroelectrics; (2) We link such topological transition to the tuning of the light beam as a ``knob'' from OAM (PS pole) to non-OAM (PS equator) modes; (3) Intermediate OAM-state structured light results in new ferroelectric topologies of temporally hybrid skyrmion-antiskyrmion states. Our study offers new approaches of robust control and flexible tuning of topologies of matter using structured light.
   
\end{abstract}

\maketitle

Originating from mathematics, the concept of topology was introduced to characterize states of matter in physics and has recently been employed over diverse physical fields~\cite{nakahara2018geometry,lin2023topological,haldane2017nobel,bernevig2022progress}. 
In real space, topological textures can manifest in various forms, including magnetic spins, electric dipoles, surface plasmon and optical fields~\cite{gobel2021beyond,junquera2023topological,shen2024optical}. Characterized by topological invariance, their intrinsic stability against external perturbations and deformations presents them as promising carriers in future information and communication technologies~\cite{parkin2008magnetic,seidel2019nanoelectronics,wan2023ultra}. Therefore, manipulation of topological structures and transitions lays basis in diverse physical systems, from optical field to materials. 

As a typical geometric representation, Poincar\'{e} sphere (PS) is used for engineering the polarization state of light~\cite{poincare1892theorie}, also extendable for orbital angular momentum (OAM) of optical vortices~\cite{padgett1999poincare}. More advanced PS models were also proposed for parametrizing complex structured light, e.g., the higher-order PS for vector beams with spatially-winding polarization pattern~\cite{milione2011higher}, Majorana sphere for highly symmetric structured beams~\cite{gutierrez2020modal}, and others for singular scalar and vector fields~\cite{yi2015hybrid,ren2015generalized,shen20202,dennis2017swings}, even plasmonic field modes~\cite{dai2022poincare}.
Moreover, it was recently topical to construct topological skyrmions by polarization vectors of light and PS-like models were also proposed to characterize optical skyrmion topologies~\cite{gao2020paraxial,shen2021topological,shen2022generation,ye2024theory}. 
By mapping topological states onto proper geometric models, light's structures can be continuously tuned and tailored, enabling the design of complex field patterns with distinct topologies~\cite{naidoo2016controlled,devlin2017arbitrary,cisowski2022colloquium,cohen2019geometric}. 

In condensed matter physics, one of the most studied topological solitonic structures is skyrmion, featured by swirling order parameters similar to a vortex on a two-dimensional (2D) lattice plane~\cite{bogdanov2020physical,das2019observation}. Thanks to its stability and emergent electrodynamics, the quasiparticle has been envisaged in high-density data storage and processing~\cite{fert2013skyrmions,han2022high}. Antiksyrmion is another prominent topological soliton, characterized by hyperbolic spin or dipole texture and is distinct from skyrmion~\cite{nayak2017magnetic}. The magnetic antiskyrmion exhibits a parallel motion on the racetrack under certain conditions, reducing data loss encountered with skyrmion~\cite{huang2017stabilization,dai2023electric}. Nevertheless, owing to its anisotropic nature, antiskyrmion is rarely discovered in ferroelectric materials~\cite{gonccalves2024antiskyrmions,gomez2024switchable}. Therefore, robust control over topological textures is highly desired, which is crucial in advancing soliton-based device applications.

The recently emerged methods of ultrafast light driving magnetic and ferroelectric topological textures~\cite{finazzi2013laser,berruto2018laser,stoica2019optical,buttner2021observation,li2021subterahertz} inspire us a new way to directly couple optical topological engineering method for characterizing topological solitons. In addition to heating effect introduced by the laser~\cite{berruto2018laser, buttner2021observation, koshibae2014creation,strungaru2022ultrafast}, the electromagnetic field of the light are also considered in governing dynamics of spins and electric dipoles and shaping their structures~\cite{yudin2017light,khoshlahni2019ultrafast,vinas2022microscopic,ghosh2023ultrafast,yambe2024dynamical,rijal2024dynamics}. Notably, with advancements in tailoring spatially nonuniform optical patterns~\cite{shen2019optical,forbes2021structured,he2022towards}, the possibility of transferring topology from structured light to condensed matter systems via light-matter interaction garners great attention~\cite{quinteiro2022interplay,habibovic2024emerging}. Several studies have demonstrated successful imprinting of optical vortices onto magnetic or ferroelectric materials~\cite{fujita2017encoding,fujita2017ultrafast,guan2023optically,gao2023dynamical,gao2024dynamical,gao2024effective,nazirkar2024manipulating,mallick2024oam}. However, a precise control and study of topological phase transition in light-induced topological structures still remains a challenge. 

In this work, we show that, by tuning the state of structured light on an OAM-based PS, we can realize a topological control of dynamical ferroelectric solitons. In particular, when light is near  the equator of OAM PS~\cite{padgett1999poincare} (that possesses a rectangularly-symmetric Hermite-Gaussian (HG) mode), a dynamical antiskyrmion can be crafted from the normal monodomain phase. Moreover, when the HG beam is gradually converted to the circularly-symmetric Laguerre-Gaussian (LG) beam~\cite{allen1992orbital}, this antiskyrmion transitions to a temporally-hybridized dynamical skyrmion-antiskyrmion, during which the relative durations of skyrmion and antiskyrmion states can be precisely controlled. Our work thus highlight that, via the interaction between structured light and ferroelectric materials, Poincar\'{e}-sphere engineering of light can act as an effective tool in manipulating ferroelectric topological structures and accomplishing topological phase transitions in materials on demand.

\begin{figure}
\includegraphics[width = 90mm]{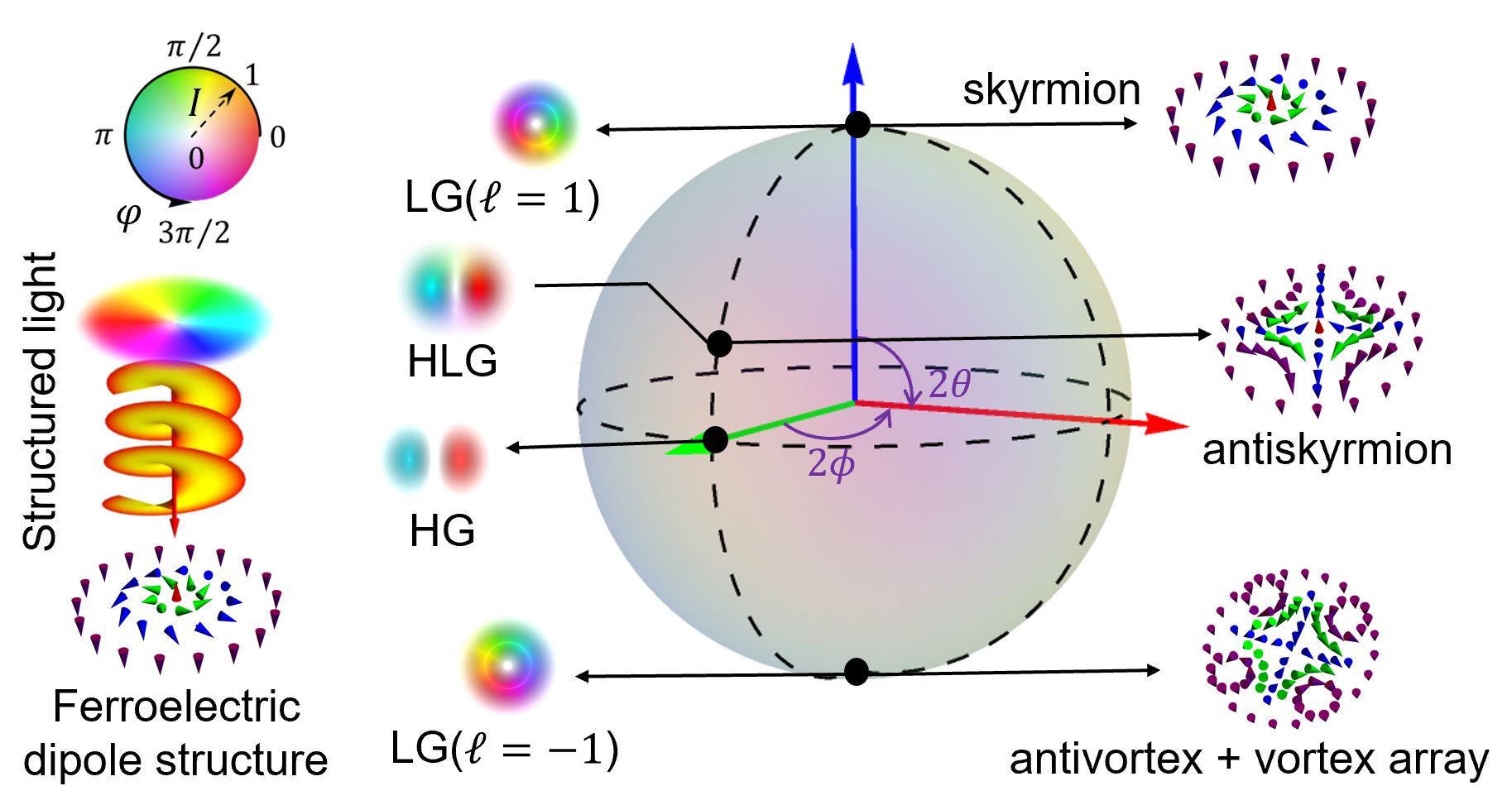}
\caption{PS engineering of structured light induced topological solitons. $2\theta$ and $2\phi$ denote the polar angle and azimuthal angle of the PS. The north and the south poles denote the LG$_{0}^{1}$ and the LG$_{0}^{-1}$ vortex beams, with the same intensity distribution but opposite phase azimuth. The equator denotes the HG$_{1,0}$ beam, with the symmetric intensity lobes but a $\pi$ phase jump between them. By varying the state of light from the north pole to the south pole along a longitudinal line, the coupled ferroelectric dipolar structure undergoes a transition from a dynamical skyrmion, to antiskyrmion, and to antivortex plus vortex array. By varying $2\phi$, the orientation of the HG beam, and that of the light-induced ferroelectric antiskyrmion, can also be adjusted. }\label{f1}
\end{figure}

In general, the spatial degrees of freedom (DoF) of light on an OAM PS can be written as:
\begin{equation}
    \ket{\Psi} = \cos{\theta} \ e^{i\phi} \ket{\ell_1} + \sin{\theta} \ e^{-i\phi} \ket{\ell_2}.
\end{equation}
As illustrated in Fig.~\ref{f1}, where $\ket{\ell}$ represent the LG$^{\ell}_{m}$ mode, $m$ and $\ell$ are radial and OAM indices, respectively. We only consider $m = 0$ in this paper for simplification. For the fundamental case, the North and South pole of the PS refer to two LG beams with opposite OAM, i.e. $\ell_1 = -\ell_2 = 1$, both characterized by a rotationally symmetric, doughnut-shape intensity distribution, and an azimuthal phase variation of $2\pi$. When $2\theta = 90^\circ$, $\ket{\Psi}$ correspond to states at the equator with equal weight of LG$_{0}^{1}$ and LG$_{0}^{-1}$. This is the first-order HG$_{1,0}$ beam, and the intensity is symmetrically distributed in two separate lobes. The intermediate state between the pole and equator is called Hermite-Laguerre-Gaussian (HLG) mode~\cite{dennis2017swings,shen20202}. Unlike LG beams, for HG beam field vectors in each lobe share the same phase, while the phases between two lobes differ by $180^{\circ}$. For the polarization DoF of the light, we choose it to be left-handed polarized, i.e., $\Vec{e} = \Vec{e}_x + i \Vec{e}_y$.

Technically, we adopt the first-principles-based effective Hamiltonian molecular dynamics to study the interaction between structured light and ferroelectric ultrathin films made of Pb(Zr$_{0.4}$Ti$_{0.6}$)O$_3$ (PZT)~\cite{zhong1994phase,zhong1995first, garcia1998electromechanical,bellaiche2001electric}, as used in our previous works~\cite{kornev2004ultrathin,lai2006electric,zhang2017nanoscale,nahas2020inverse,nahas2020topology,prokhorenko2024motion}. All computational details and related discussions are presented in the Supplementary Material (SM). Here, we briefly mention that the initial dipole configuration is a polar monodomain with the polarization oriented along [00$\bar{1}$] direction, and the structured light is normally incident on the $xy$ plane so that its electric field can be directly coupled with the in-plane dipole components $p_{x,y}$. The light frequency $\omega$ is chosen as 1 THz for driving the motion of ions, and the beam radius is set to 5 unit cells (u.c.), compatible with the dimension of the simulation cell ($80\times 80 \times 5$ u.c.). Note that recent developments in plasmonics and near-field techniques can help to overcome the diffraction limit~\cite{novotny2011antennas,gramotnev2014nanofocusing,heeres2014subwavelength,prinz2023orbital}; also in principle, the beam radius can be further increased without changing the physics. As illustrated in Fig. 1, by varying the state of the light along a longitude from the South pole to the North pole on the OAM PS, the induced ferroelectric topological structure undergoes corresponding changes, which are discussed in details below.

\begin{figure}
\includegraphics[width = 80mm]{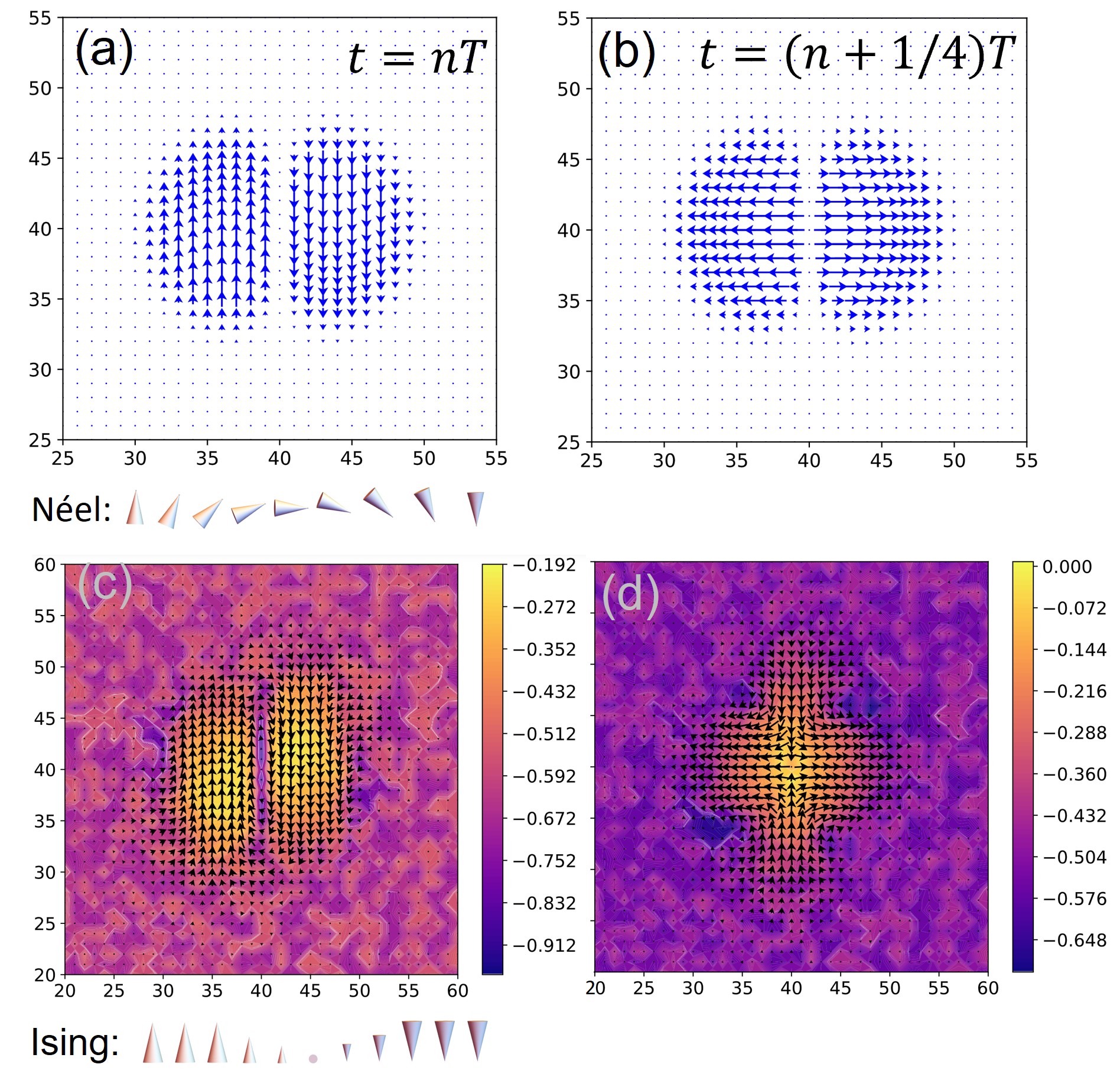}
\caption{ Distribution of electric field vectors of the HG$_{1,0}$ beam at times (a) $t=nT$ and (b)$t=(n+1/4)T$ on the (001) ($xy$) plane. Accordingly, the induced dipole pattern on the bottom (001) layer right after $t=nT$ and $t=(n+1/4)T$  are presented in (c) and (d). $x$ and $y$ axis denote the unit cell number along the respective in-plane direction. In-plane dipole components $p_x$ and $p_y$ are denoted by the vector, and out-of-plane dipole components $p_z$ are denoted by the colorbar. The Ising-type and N\'{e}el-type domain walls, which correspond to $p_{x,y}$ along $x$ direction in $35 \le y \le 45$ and $5 < |y - 40| < 10$, are illustrated on the bottom and top part of panel (c).  }\label{f2}
\end{figure}

Figures 2(a)(b) present the electric field vector distribution $\bm{E}(\bm{r})$ of the HG$_{1,0}$ beam---represented by $\phi = 0, 2\theta = 90^\circ$ on the PS---on the (001) ($xy$) plane of the thin film, at times $t = nT$ and $t = (n + 1/4) T$, with $n$ being an integer and $T$ being the period of the light. The fields in the two lobes are antiparallel to each other, and in each lobe, the field is maximized at the lobe center and gradually decreases towards the edge. The angle $\phi = 0 $ determines that the symmetry axis, which also serves as the boundary between two lobes, positioned at $x = 40$. With the  circular polarization of light, a phase difference of $90^{\circ}$ is accumulated over a quarter of period $T/4$, resulting in fields aligned along horizontal and vertical directions in Fig.~\ref{f2}(a) and Fig.~\ref{f2}(b). The field evolution within a whole period $T$ is presented in the animation in SM. Note that if the electric dipole $\bm{p}(\bm{r})$ rigorously follows the on-site field $\bm{E}(\bm{r})$, the transition region from one lobe to the other across $x = 40$ corresponds to an Ising-type and a tail-to-tail domain wall with respect to Fig.~\ref{f2}(a) and  Fig.~\ref{f2}(b) ~\cite{nataf2020domain,hong2021vortex}. As to be seen, this is closely connected to the emergent dipolar textures following the structured light. 

Figure~\ref{f2}(c) presents the dipole pattern formed on the bottom layer of the film in response to the electric field of Fig.~\ref{f2}(a) at time $t = nT$. For $ 35 \le y \le 45$, strong local $\bm{E}(\bm{r})$ induces significant and antiparallel $p_y$ within two lobes owing to the $\bm{p}(\bm{r}) \cdot \bm
{E} (\bm{r})$ coupling, analogous to the Zeeman coupling in the magnetic counterpart~\cite{fujita2017ultrafast,khoshlahni2019ultrafast,guan2023optically,lei2024rotational}. At the lobe boundary $x = 40$, the Ising-type domain wall characterized by negligible $p_x$ and $p_y$ in this region is preserved for the in-plane dipole components. Such domain wall is charge-neutral, and its formation does not incur a high energy cost~\cite{bednyakov2018physics}. Meanwhile, the out-of-plane dipole components $p_z$ become dominant to preserve the continuity of the dipole magnitude. In contrast, in the region $5 < |y - 40| < 10$, $p_x$ and $p_y$ adjust their orientations more freely under a small local  $\bm{E}(\bm{r})$. As a result, a N\'{e}el-type domain wall across two lobes along $x$ direction with continuously rotated $p_{x,y}$ is observed. Overall, an elliptical dipolar vortex ring forms on the (001) plane, consistent with flux-closure or vortex domains discovered in PZT systems that compensate for the polar discontinuity at the domain boundary~\cite{jia2011direct,wang2014origin,wei2016neel}. 

Even more intriguing is the dipole response to the electric field of Fig.~\ref{f2}(b). Figure~\ref{f2}(d) presents the dipole pattern on the bottom layer right after $t=(n+1/4)T$, and the in-plane dipole components form an antivortex. Specifically, dipoles in the horizontal ``wing" region ($ 3 \le | x - 40| \le 7$, $36 \le y \le 44 $) exhibit minor $p_z$ and major $p_x$ components pointing away from the center and align with the local horizontal electric field; when they move closer toward the center along $x$ direction, dipoles at the edges of the wing gradually reorient along $y$ direction; this gradually transitions to the vertical ``wing" region ($ 37 \le x \le 43$, $ 5 \le | y - 40| \le 9$), where $p_{x,y}$ primarily align along $y$ direction pointing toward the center. Interestingly, with tiny local $\bm{E}(\bm{r})$ in the vertical ``wing" region, the development of $p_y$ should be regarded as a nonlocal response, originating from dipole-dipole interaction and depolarization effect rather than the local $\bm{p}(\bm{r}) \cdot \bm
{E} (\bm{r})$ coupling~\cite{ponomareva2005atomistic,ponomareva2005low}. Note that in both Figs. 2(c) and (d), $p_z$ remains negative throughout the plane, indicating that no {\it topological} phase transition occurs between the initial down-poled monodomain and the formed vortex ring or antivortex. 

\begin{figure}
\includegraphics[width = 90mm]{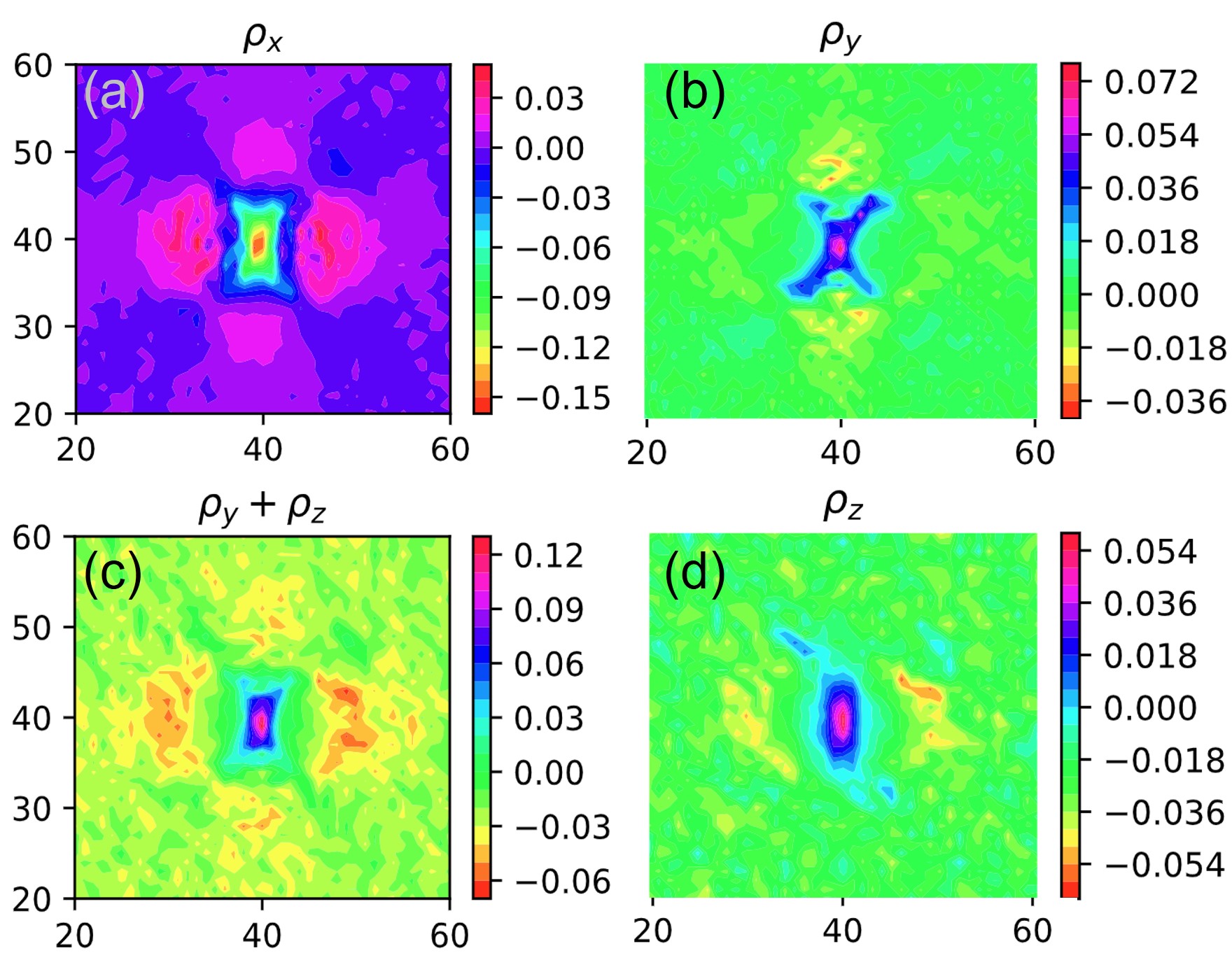}
\caption{Distribution of bound charge components (a) $\rho_{{\rm{b}},x}$, (b) $\rho_{{\rm{b}},y}$, (c) $\rho_{{\rm{b}},y} + \rho_{{\rm{b}},z}$ and (d) $\rho_{{\rm{b}},z}$, associated with the dipole pattern presented in Fig.~\ref{f2}(d). $x$ and $y$ axis denote the unit cell number along the respective in-plane direction. The colorbar denotes the charge value in an arbitrary unit. }\label{f3}
\end{figure}

The polar antivortex in Fig.~\ref{f2}(d) with notable $p_y$ is different from electric fields in Fig.~\ref{f2}(b), which only have $E_x$ components. To elucidate the origin of the antivortex formation, we plot the distribution of the electric bound charge density $\rho_{{\rm{b}}}(\bm{r}) = - \nabla \cdot \bm{P}(\bm{r})$, which arises from the spatially nonuniform polarization. In general, a large bound charge can greatly increase the electrostatic energy, causing instability in the domain~\cite{hlinka2019skyrmions,chen2021recent,junquera2023topological}; such instability can be mitigated with the self-organized polarization pattern bearing minimal divergence, effectively neutralizing $\rho_{{\rm{b}}}$~\cite{luk2020hopfions,tikhonov2022polarization}. Figure~\ref{f3}(a) presents the bound charge component $\rho_{{\rm{b}},x}$ contributed by $p_x$ at $t=(n+1/4)T$, manifesting as large negative values at the layer center. The pronounced $\rho_{{\rm{b}},x}$ is associated with the intrinsically unstable tail-to-tail $p_x$ pattern dictated by $E_x$, whereas the presence of such pattern in certain ferroelectrics has been attributed to the external screening effect from mobile charge defects or free carriers~\cite{bednyakov2016free,bednyakov2018physics}. Here, in multiaxial ferroelectric PZT, the rotational flexibility of the polarization enables the reduction of $\rho_{{\rm{b}},x}$  through the synergistic action of $p_y$ and $p_z$: The $p_y$s along the lobe boundary at $x = 40$ resemble a head-to-head domain wall, and form the antivortex together with the $p_x$; meanwhile, at the central region in each layer, magnitudes of the negative $p_z$ continuously decrease from the top to the bottom layer (see section view in Fig. S1 in SM). Figures~\ref{f3}(b) and (d) present the bound charge components $\rho_{{\rm{b}},y}$ and $\rho_{{\rm{b}},z}$, both of which demonstrating positive values at the layer center. When they are added together, they effectively counteract $\rho_{{\rm{b}},x}$, as shown in Fig.~\ref{f3}(c). The mechanism that in-plane and out-of-plane dipole components orthogonal to the one-dimensional (1D) leading electric field respond collectively to minimize the net bound charge is also observed when the leading field and antivortex are rotated with $2\phi$ (see Fig. S3 in SM). This differs from the scenario where the leading electric field is 2D, e.g., in the shape of a vortex, which dictates both $p_x$ and $p_y$, while only $p_z$ act as the responsive component~\cite{gao2024dynamical}.    

\begin{figure}
\includegraphics[width = 80mm]{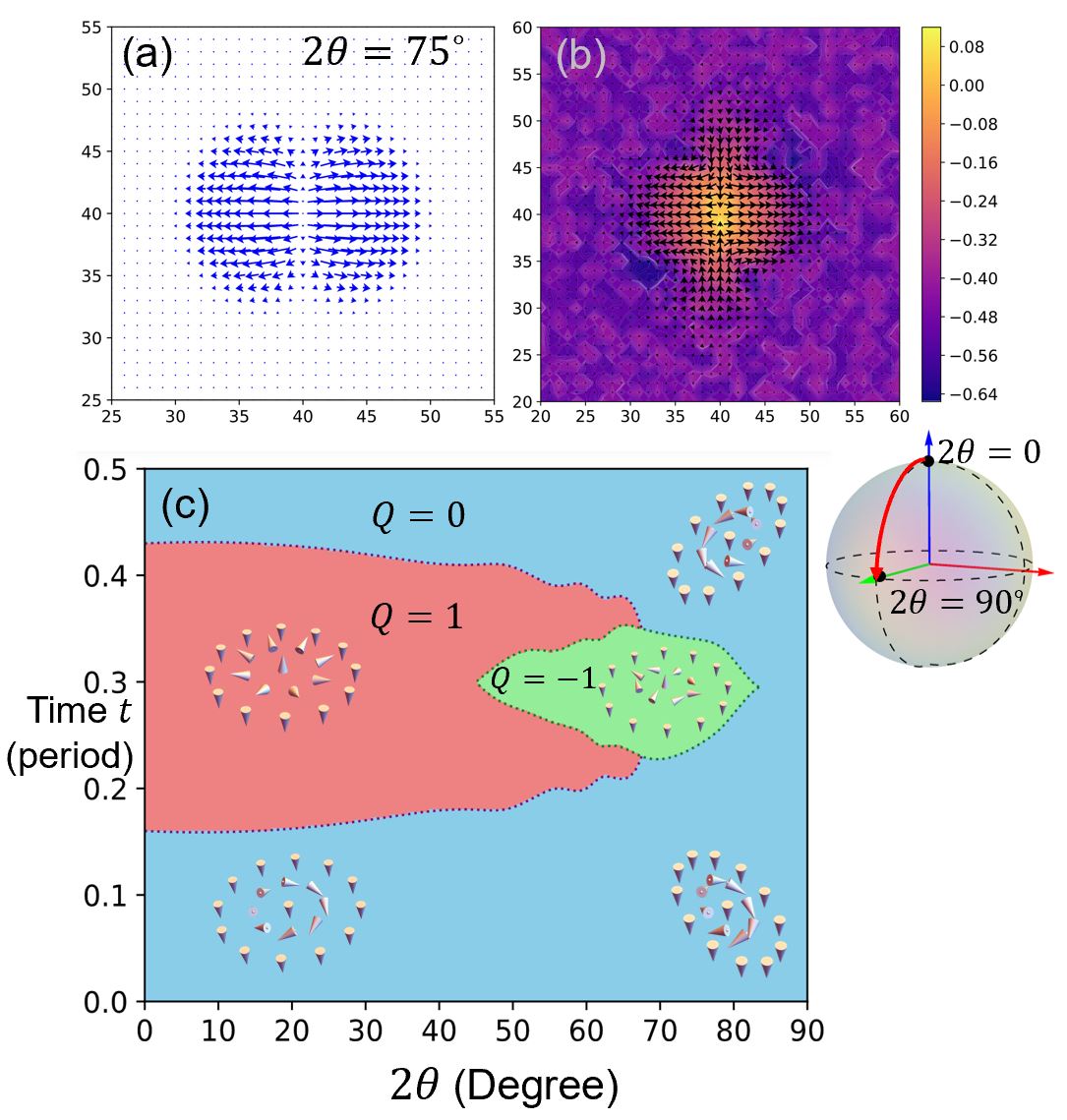}
\caption{Distribution of (a) electric field vectors and (b) the induced dipole pattern on the (001) ($xy$) plane at time $t=(n+1/4)T$ when the state of the light corresponds to $2\theta = 75^{\circ}, 2\phi = 0$ on the OAM PS. (c) The established phase diagram of the dipole configuration on a given surface layer by varying the $2\theta$ angle of the OAM PS and the time $t$. }\label{f4}
\end{figure}

So far, with negative $p_z$ across the plane, we have not yet achieved a topologically distinct phase. However, as shown in Fig.~\ref{f2}(c), magnitudes of $p_z$ at the central region of the bottom layer are close to zero, granting an opportunity to generate an antiskyrmion if the central $p_z$ can be flipped (i.e., become positive). A flipped $p_z$ at the bottom layer reflects a greater difference compared to $p_z$ at the adjacent upper layer, indicating a larger local $\rho_{{\rm{b}},z}$. Considering the synergy between $p_y$ and $p_z$ (or $\rho_{{\rm{b}},y}$ and $\rho_{{\rm{b}},z}$) to counteract $\rho_{{\rm{b}},x}$, this raises the question of whether we can reduce $|\rho_{{\rm{b}},y}|$ to increase $|\rho_{{\rm{b}},z}|$ and assist the flipping $p_z$. Fortunately, this can be achieved by modifying the electric field utilizing OAM PS: moving from equator to the northern hemisphere, the LG beam in the form of a divergent vortex is added as an additional component of the electric field. Figure~\ref{f4}(a) show the electric field  at $t=(n+1/4)T$ when $2\theta = 75^{\circ}, 2\phi = 0^{\circ}$. In such case, the centrifugal $E_y$ oppose the centripetal $p_y$, reducing their magnitudes and associated $|\rho_{{\rm{b}},y}|$ (see Fig. S4 in SM for the respective $\rho_{{\rm{b}}}$). As a result, $p_z$ in the central region now exhibit positive values, and $p_{x,y}$ in the form of an antivortex are still maintained, resulting in the formation of an antiskyrmion (see Fig.~\ref{f4}(b))!  Note that this antiskyrmion therefore does not really form at the equator but rather a little bit tilted towards the north from it. The antiskyrmion can be rigorously characterized by computing the topological charge $Q$ through integrating the Pontryagin charge density $\rho_{\rm{sk}}(\bm{r})$ over the full plane:
\begin{equation}
    Q = \int \rho_{\rm{sk}}(\bm{r}) d^2 \bm{r} =  \int \frac{1}{4\pi}\bm{n}(\bm{r})\cdot\big( \frac{\partial \bm{n}(\bm{r})}{\partial x} \times \frac{\partial \bm{n}(\bm{r})}{\partial y} \big) d^2 \bm{r},
    \label{s10}
\end{equation}
where $\bm{n}(\bm{r})$ is the unit dipole field vector at site $\bm{r}$. The notable negative part of the topological charge density $\rho_{\rm{sk}}(\bm{r})$ is concentrated in the central region (see Fig. S5 in SM). Consistent with the observation, the corresponding $Q$ is indeed $-1$. Note that our previous work revealed that such polar antiskyrmion cannot be produced by an LG beam in the shape of an antivortex (i.e., the South pole of the OAM PS)~\cite{gao2024effective}, which can only induce an antivortex plus vortex array, as illustrated in Fig. 1. 

Since the electric field vectors evolve periodically over time and exhibit distinct patterns at different times, we construct a phase diagram to characterize the topological phase on the surface layer as a function of time $t$ and $2\theta$ angle in Fig.~\ref{f4}(c). Azimuthal angle $2\phi$ is not included since it only changes the orientation but not the topology of the emergent structure (see Fig. S3 and discussion in SM). Topological charge at respective time and $2\theta$ angle is calculated to differentiate each phase. Note that only one half of the period $T$ on the $y$ axis is shown, as the other half is topological trivial ($Q = 0$) for the given surface layer, but nontrivial for the other surface layer due to symmetry (see Fig. S1, S2 and discussion in SM). The southern hemisphere of OAM PS ($2\theta > 90^{\circ}$) is also not shown as the phase is always topological trivial ($Q = 0$). At $2\theta = 90^{\circ}$, the dipolar structure evolves from a clockwise, elliptical vortex ring, to an antivortex, and finally to an anticlockwise vortex ring with time within the duration of a half period. As $2\theta$ decreases to $82.5^{\circ}$, the antivortex emerges around $t = 1/4 \  T$, and its timespan further extends with the increasing $2\theta$. Such trend continues until $2\theta = 66^\circ$ when the skyrmion phase begins to emerge, demonstrating that centripetal $p_y$ are converted to centrifugal $p_y$ at this critical angle. A greater $2\theta$ with a larger weight of LG beam further extends the timespan of the skyrmion while shortening that of the antiskyrmion. Thus, by mixing LG and HG beams, a robust, temporally hybridized antiskyrmion-skyrmion phase can be established. When $2\theta \le 45^{\circ}$, the antiskyrmion phase vanishes completely, while the skyrmion phase persists up to $T/4$, transitioning between a clockwise and an anticlockwise vortex ring. This shows that by engineering the state of light on the OAM PS with a tuning on $2\theta$, a conversion between solitons with opposite topological charges can be realized.   

As a conclusion, we use first-principles-based calculations and demonstrate the critical role of OAM PS in the interaction between structured light and ferroelectrics, which enables the creation of dynamical antiskyrmion---a rarely discovered soliton in ferroelectric materials. The OAM PS provides a robust control over topology of the electric dipole textures, allowing for the conversion from a dynamical antiskyrmion to skyrmion. Derived from light-matter interaction, our work introduces a novel strategy in generating and controlling topological dipolar textures via engineering the OAM PS state of light. Such manipulation is in contrast to conventional topological engineering in ferroelectrics based on electrical and mechanical methods~\cite{chen2020atomic,zhu2022dynamics}. Finally, it is a promising direction to apply more complex structured light PS models for other exciting nontrivial topologies in diverse condensed matter materials.   

{\em Acknowledgement}---We acknowledge the support from the Grant MURI ETHOS W911NF-21-2-0162 from Army Research Office (ARO), the Vannevar Bush Faculty Fellowship (VBFF) Grant No. N00014-20-1-2834 from the Department of Defense, Singapore Ministry of Education (MOE) AcRF Tier 1 grant (RG157/23), MoE AcRF Tier 1 Thematic grant (RT11/23), and Imperial-Nanyang Technological University Collaboration Fund
(INCF-2024-007). We also acknowledge the computational support from Naval HPCMP Pathfinder Award and the Arkansas High Performance Computing Center.
 
\bibliography{reference.bib}



\section*{Supplementary material}

# Poincaré sphere engineering of
# dynamical ferroelectric topological solitons

Lingyuan Gao,[1, *] Yijie Shen,[2, 3, †] Sergei Prokhorenko,[1]

Yousra Nahas,[1] and Laurent Bellaiche[1, 4, ‡]

[1]*Smart Ferroic Materials Center, Physics Department*

*and Institute for Nanoscience and Engineering,*

*University of Arkansas, Fayetteville, Arkansas, 72701, USA*

[2]*Centre for Disruptive Photonic Technologies,*

*School of Physical and Mathematical Sciences,*

*Nanyang Technological University, Singapore 637371, Singapore*

[3]*School of Electrical and Electronic Engineering,*

*Nanyang Technological University, Singapore 639798, Singapore*

[4]*Department of Materials Science and Engineering,*

*Tel Aviv University, Ramat Aviv, Tel Aviv 6997801, Israel*

(Dated: February 11, 2025)

---

* These authors contributed equally

† These authors contributed equally; yijie.shen@ntu.edu.sg

‡ laurent@uark.edu

# I. COMPUTATIONAL METHODS AND DETAILS

Pb(Zr$_{0.4}$Ti$_{0.6}$)O$_3$ (PZT) ferroelectric ultrathin films grown along the [001] direction is modeled using first-principles-based effective Hamiltonian approach[1–4], a successful method in predicting unconventional phase transitions in this system[5–9]. For this effective Hamiltonian, different types of interactions involving dipoles and elastic strains are included in the total energy, described by local B-centered soft modes in each unit cell (u.c.) and strain variables, respectively. We use an $80 \times 80 \times 5$ simulation supercell and apply a $\sim$-2% compressive strain along the periodic [100] and [010] pseudo-cubic directions. After gradually cooling the system with a similar strategy that adopted in Refs. [ 9,10], we perform molecular dynamics at 10 K at a resolution of 0.5 femtoseconds (fs) on the well-thermalized monodomain configuration with all electric dipoles aligned along the [00$\bar{1}$] direction. To model light-matter interaction, we add the Zeeman-like coupling term as $-\sum_i \boldsymbol{E}_i \cdot \boldsymbol{p}_i$, where $\boldsymbol{E}_i$ and $\boldsymbol{p}_i$ refer to the local electric field and dipole moment in cell $i$[11,12]. It is important to note that in addition to the PZT effective Hamiltonian[3,4], a term mimicking the screening of the depolarization field inside the ferroelectric nanostructures should also be included[13,14]: $\beta = 0$ and $\beta = 1$ represent the open-circuit and short-circuit conditions, which correspond to no-screening and full-screening of the depolarization field, respectively. In this work, we choose $\beta$ to be 0.87, which has been previously proven to be a realistic value[9,10,15].

In the current study, the north and the south pole of the Poincaré sphere are left-handed Laguerre Gaussian modes LG$_{m=0}^{\ell=1}$ and LG$_{m=0}^{\ell=-1}$. In the cylindrical coordinate $(r, \varphi)$, their electric fields are given by:

$$\vec{E}_{m=0, \ell=\pm 1}(t) = E_0 \sqrt{\frac{2}{\pi}} (\frac{\sqrt{2}r}{w}) e^{-\frac{r^2}{w^2}} \operatorname{Re}\{e^{i\ell\varphi + i\omega t}\vec{e}_x + e^{i(\ell\varphi - \pi/2) + i\omega t}\vec{e}_y\}. \tag{S1}$$

We set $\omega = 1$ Terahertz (THz) as the field frequency, compatible with the response time of electric dipoles. We choose $w = 5$ u.c. as the beam radius of the field—recent advancements in nanoscale light focusing far below the free-space diffraction limit makes such optical field promising[16–20]. $E_0$ is set to match a magnitude of $\sim 10^8$ V/m in experiment[21,22], which does not cause a dielectric breakdown due to the significantly reduced dielectric constant in the THz frequency range[23–25].

## II. TIME-EVOLVING ELECTRIC FIELDS OF POINCARÉ- SPHERE-REPRSENTED MODES AND INDUCED DIPOLE CONFIGURATIONS

Figures S1 present the time-evolving electric field vector distribution of the $HG_{1,0}$ beam (represented by the equator $2\phi = 0, 2\theta = \pi/2$ on the Poincaré sphere) and the following dipole configurations from different views at different quarters of a period $T$. At $t = nT$ and $(n + 1/2)T$, section views in Fig. S1(b1)/(b3) reveal that $p_z$ behaves more uniformly across layers in the absence of centrifugal/centripetal electric field components (Fig. S1(a1)/(a3)); as a result, elliptical vortices form on both the top and bottom layers, as shown in the in-plane views Fig. S1(c1)/(c3)/(d1)/(d3). In contrast, at $t = (n + 1/4)T$ and $(n + 3/4)T$, when the electric field points toward or away from the center (Fig. S1(a2)/(a4)), $p_z$ exhibits notable variations from the top to the bottom layer (Fig. S1(b2)/ (b4)). Depending on the electric field direction, the homogeneous antivortex (Fig. S1(c2)/ (d4)) emerges on either the bottom or top layer, accompanied by small $p_z$ located in the central region of the same layer. Such distribution can be explained by the bound charge density shown in the next section. The whole period of electric field evolution and in-plane view of the dipole configuration on the top layer are presented as separate animations (Field-2theta90.gif and Dipole-2theta90-top.gif).

Similarly, Figures S2 presents the time-evolving electric field vector distribution of the $LG_{m=0}^{\ell=-1}$ beam (represented by the South pole $2\phi = 0, 2\theta = \pi$ on the Poincaré sphere) and the following dipole configurations at different times. At all four quarters, the section views (Fig. S2(b1)-(b4)) reveal that $p_z$ remains uniform across layers, while the in-plane views (Fig. S2(c1)-(c4), (d1)-(d4)) show the formation of an antivortex along with a vortex array of four vortices on both the top and bottom layers. Consistent with the observation in Ref. 26, the antivortex plus vortex array gyrate over time. For $LG_{m=0}^{\ell=1}$ beam (represented by the north pole $2\phi = 0, 2\theta = 0$ on the Poincaré sphere), the electric field and associated dipole configurations are not presented here and are referred to Ref. 27.

We also present the electric field and associated dipole configurations of the 45°-rotated $HG_{1,0}$ beam (represented by the equator $2\phi = 3\pi/2, 2\theta = \pi/2$ on the Poincaré sphere) at $t = (n + 1/4)T$ in Fig. S3(a)-(d). Compared to Fig. S1(a1)-(d1), an in-plane rotation controlled by $\phi$ is applied on the electric field and in-plane dipole components.

## III. ELECTRIC BOUND CHARGE DENSITY AND TOPOLOGICAL CHARGE DENSITY

In Fig. 3 of the main manuscript, we show that the electric bound charge components $\rho_{\mathrm{b},y}$ and $\rho_{\mathrm{b},z}$ work synergistically to counteract $\rho_{\mathrm{b},x}$. To support this argument, we plot corresponding bound charge components for rotated dipole configurations on the bottom and the top layers presented in Fig. S3(c)(d), and show them in Fig. S3(e)-(l). Consistent with Fig. 3 in the main manuscript, here in the 45°-rotated configuration, when $\rho_{\mathrm{b},y'}$ and $\rho_{\mathrm{b},z}$ are added together, they effectively compensate $\rho_{\mathrm{b},x'}$.

In Fig. S4, we compare bound charge components on the bottom layer at $t = (n + 1/4)T$ between $2\theta = 90°$ (Fig. S4(a)-(d)) and $2\theta = 75°$ (Fig. S4(e)-(h)). Indeed, at $2\theta = 75°$, introducing the divergent vortex beam as an additional electric field component causes $\rho_{\mathrm{b},y}$ to decrease while $\rho_{\mathrm{b},z}$ to increase, resulting in flipped, positive $p_z$ in the central region and the formation of antivortex.

In Fig. S5, we present the Pontryagin charge density $\rho_{\mathrm{sk}}(\boldsymbol{r})$ for antiskyrmion, Néel-type skyrmion and "Bloch-like" skyrmion. Integration of the Pontryagin charge density yields the topological charge $Q = -1, +1$ for antiskyrmion and skyrmion, respectively. As shown in Fig. S5, nonzero $\rho_{\mathrm{sk}}$ are concentrated in the central region of the layer.

## IV. ANIMATIONS OF TUNABLE DYNAMICAL FERROELECTRIC TOPOLOGICAL SOLITONS

We present animations depicting the evolution of the electric field and dipole configuration on the top layer for $2\theta = 75°$ and $2\theta = 60°$ on Poincaré sphere. In the animations, $Q$ represents the topological charge. For $2\theta = 75°$, the transition from a trivial phase → antiskyrmion → a trivial phase is shown (Field-2theta75.gif and Dipole-2theta75-top.gif). For $2\theta = 60°$, the transition from a trivial phase → skyrmion → antiskyrmion → skyrmion → a trivial phase is shown ((Field-2theta60.gif and Dipole-2theta60-top.gif)).

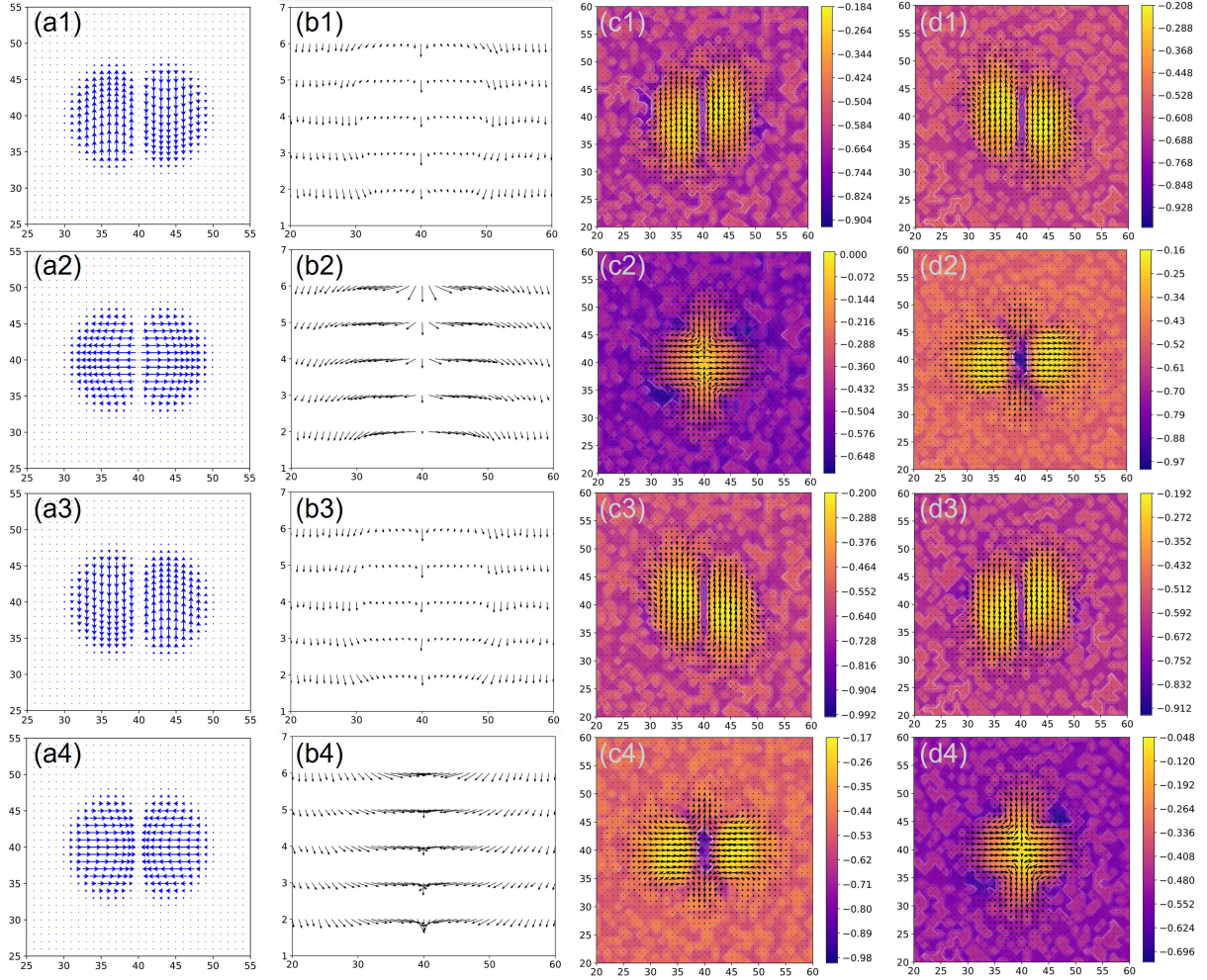

FIG. S1. Electric field vector distribution of the $HG_{1,0}$ beam and the induced dipole configurations from different views over times. Rows 1–4 correspond to times from $t = nT$ to $t = (n + 3/4)T$ at an interval of $T/4$. Columns (a)-(d) represent (a) the field distribution on the (001) ($xy$) plane, (b) a section view of the induced dipole configuration on the (010) ($xz$) plane showing $p_x$ and $p_z$ components, (c) an in-plane view ($xy$ plane) of the dipole configuration at the bottom layer, with $p_x$ and $p_y$ components denoted by the vector, and $p_z$ component denoted by the colorbar, and (d) an in-plane view of the dipole configuration at the top layer.

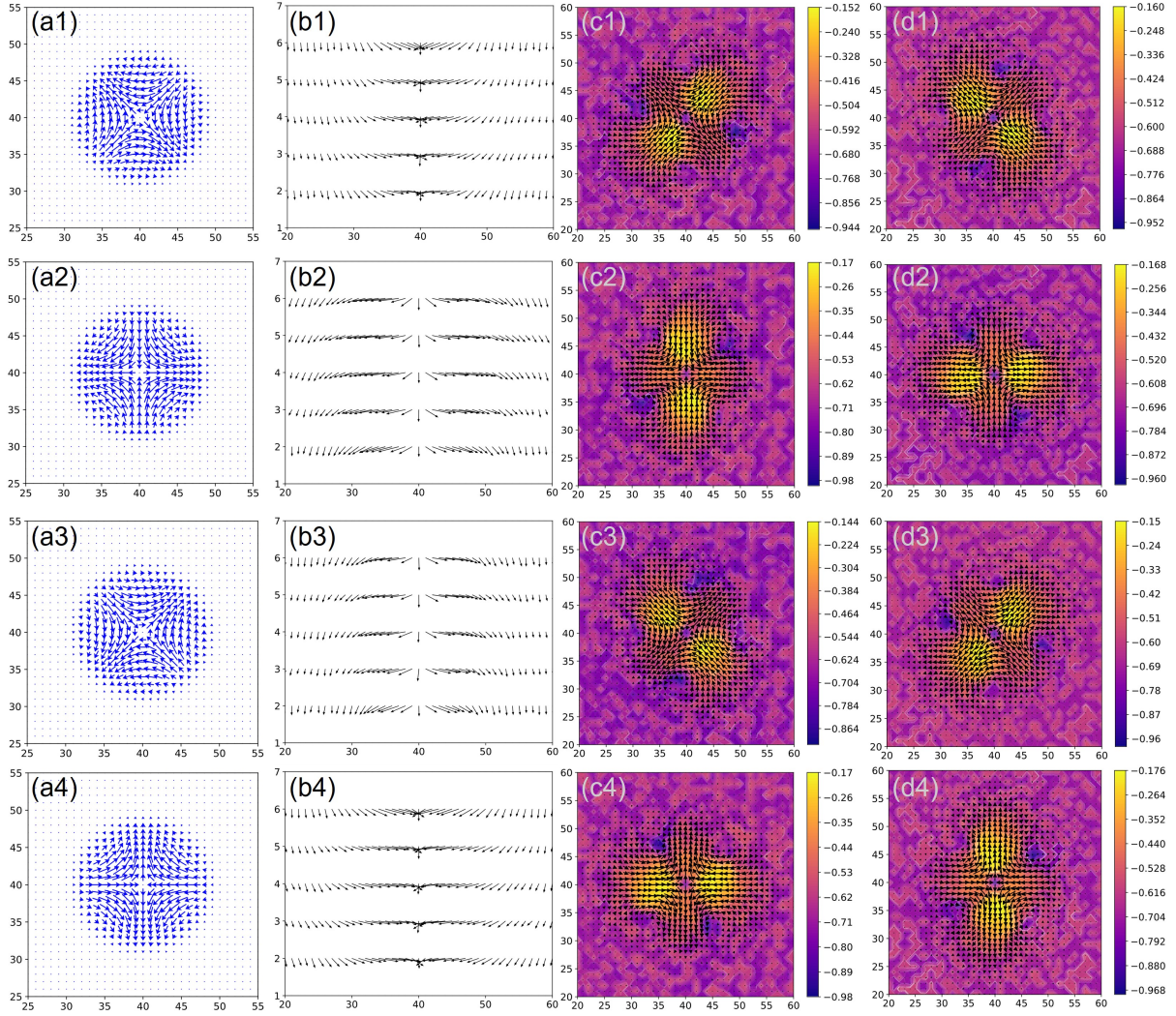

FIG. S2. Electric field vector distribution of the $\mathrm{LG}_{m=0}^{\ell=-1}$ beam and the induced dipole configurations from different views over times. Row 1–4 correspond to times from $t = nT$ to $t = (n+3/4)T$ at an interval of $T/4$. Columns (a)-(d) represent (a) the field distribution on the $(001)$ $(xy)$ plane, (c) an in-plane view of the dipole configuration at the bottom layer, and (d) an in-plane view of the dipole configuration at the top layer. Column (b) presents the section view of the induced dipole configuration: (b1) and (b3) show the view from $(\bar{1}10)$ plane, displaying $(p_x + p_y)/\sqrt{2}$ and $p_z$ components; (b2) and (b4) show the view from $(010)$ plane, displaying $p_x$ and $p_z$ components.

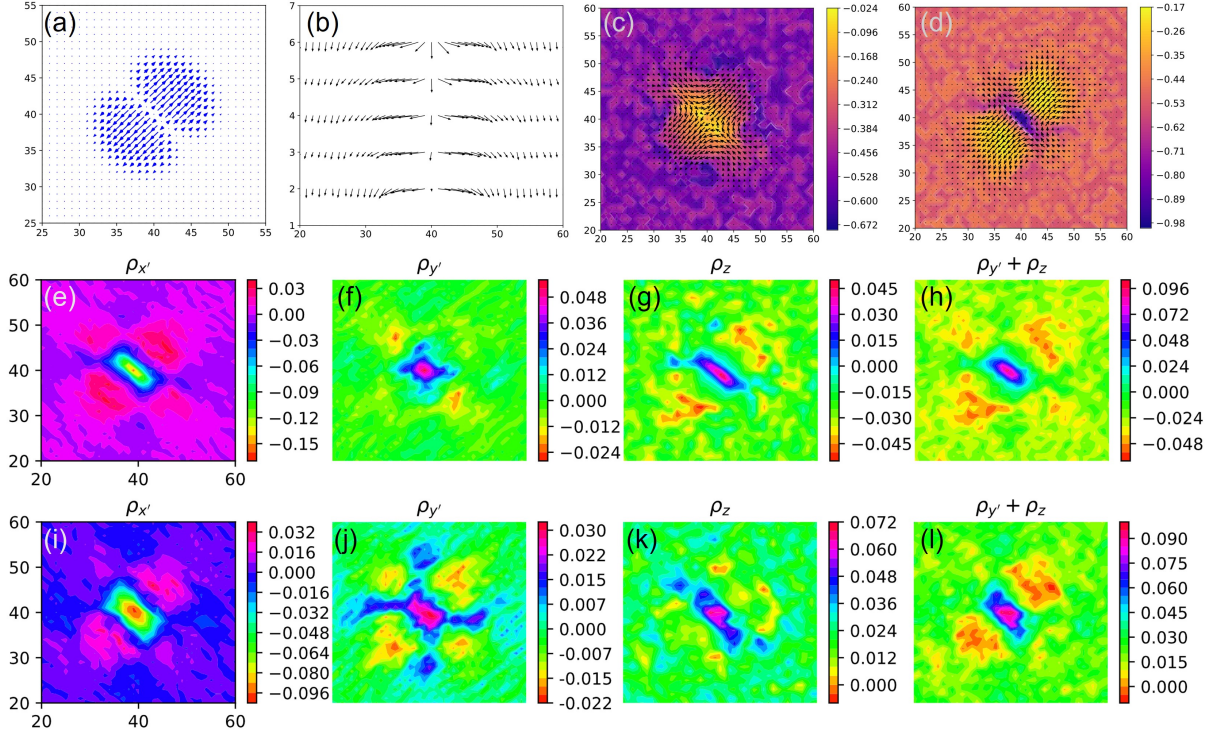

FIG. S3. (a)-(d) Similar to Fig. S1 and Fig. S2, electric field vector distribution of the 45°-rotated $HG_{1,0}$ beam and the induced dipole configurations from different views at time $t = (n + 1/4)T$. (e)-(h) Distribution of electric bound charge components $\rho_{b,x'}$, $\rho_{b,y'}$, $\rho_{b,z}$ and $\rho_{b,y'} + \rho_{b,z}$ on the bottom layer in the rotated coordinate, with $\boldsymbol{x'} = \frac{1}{\sqrt{2}}(\boldsymbol{x} + \boldsymbol{y})$, $\boldsymbol{y'} = \frac{1}{\sqrt{2}}(\boldsymbol{x} - \boldsymbol{y})$. (i)-(l) Similar to (e)-(h), but on the top layer.

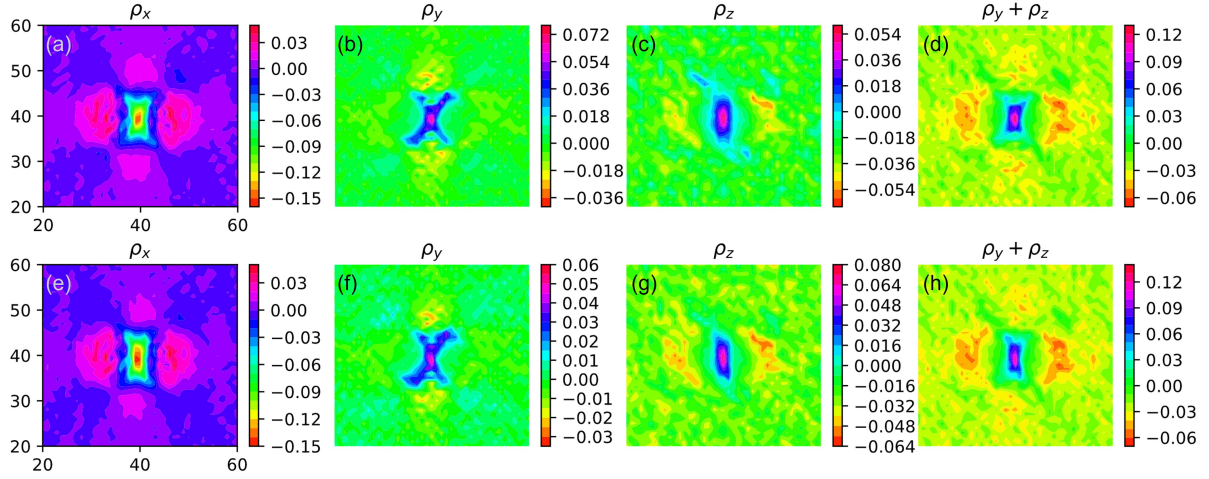

FIG. S4. (a)-(d) Distribution of electric bound charge components on the bottom layer for $HG_{1,0}$ beam ($2\phi = 0, 2\theta = \pi/2$ on the Poincaré sphere) at time $t = (n + 1/4)T$. (e)-(f) Similar to (a)-(d), but for HLG beam ($2\phi = 0, 2\theta = 75°$ on the Poincaré sphere).

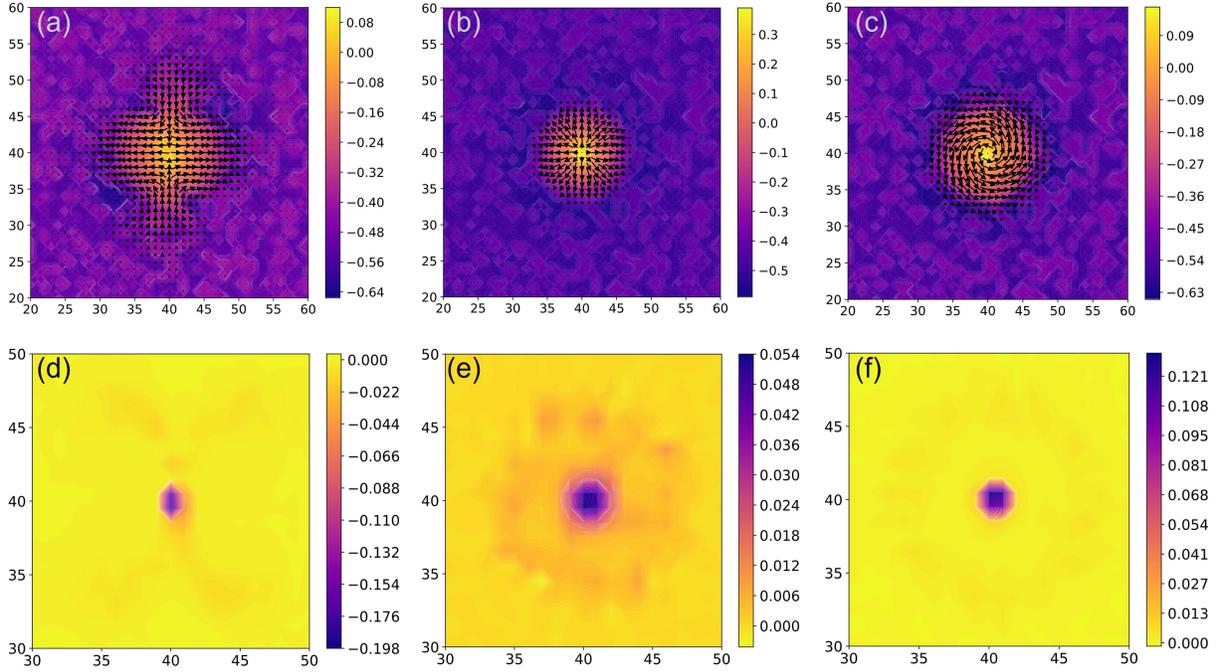

FIG. S5. In-plane views of (a) antiskyrmion, (b) Néel-type skyrmion, and (c) Bloch-type skyrmion, along with (d)-(f) their corresponding Pontryagin charge density distributions $\rho_{sk}(\boldsymbol{r})$.



\end{document}